%% file: main.tex
\documentclass[preprint,12pt]{elsarticle}
\usepackage[T1]{fontenc}
\usepackage{lmodern}
\usepackage{amsmath,amssymb}
\usepackage{graphicx}
\usepackage{booktabs}
\usepackage{flafter}

\usepackage{siunitx}
\usepackage{xcolor}
\usepackage{orcidlink}
\AtBeginDocument{\hypersetup{hidelinks}}

\newcommand{\VMn}{$V_{\mathrm{Mn}}$}
\newcommand{\VSe}{$V_{\mathrm{Se}}$}
\newcommand{\SeMn}{$\mathrm{Se}_{\mathrm{Mn}}$}
\newcommand{\MnSe}{$\mathrm{Mn}_{\mathrm{Se}}$}
\newcommand{\dmu}{\Delta\mu_{\mathrm{Se}}}
\journal{Physica B: Condensed Matter}

\begin{document}
\begin{frontmatter}

\title{Native vacancies and antisites in rock-salt and wurtzite MnSe: formation energies across the
growth window and spin splitting in the A-type order}

\author[cicima,ucr]{Tom\'as Rojas Solorzano\,\orcidlink{0000-0001-5745-4255}\corref{cor1}}
\ead{tomas.rojas\_s@ucr.ac.cr}
\cortext[cor1]{Corresponding author}
\author[umd]{Dale Igram\,\orcidlink{0000-0003-4450-1483}}

\affiliation[cicima]{organization={Centro de Investigaci\'on en Ciencia e Ingenier\'ia de
            Materiales (CICIMA), Universidad de Costa Rica},
            postcode={11501-2060},
            city={San Jos\'e},
            country={Costa Rica}}
\affiliation[ucr]{organization={Sede de Occidente, Universidad de Costa Rica},
            postcode={20201},
            city={San Ram\'on},
            state={Alajuela},
            country={Costa Rica}}
\affiliation[umd]{organization={University of Michigan--Dearborn},
            addressline={4901 Evergreen Road},
            postcode={48128},
            city={Dearborn},
            state={Michigan},
            country={USA}}

\begin{abstract}
Manganese selenide grows both as rock-salt $\alpha$-MnSe, a type-II antiferromagnet, and as
metastable wurtzite w-MnSe, an antiferromagnet whose type of order is not known. In DFT+$U$, a
spin-degenerate stripe order of w-MnSe lies 14~meV per formula unit below the A-type order, which is
predicted to be altermagnetic. We compute the neutral Mn and Se vacancies and the two antisites
of both polymorphs with a linear-response Hubbard $U$, between the Se-rich limit bounded by MnSe$_2$ and
the Mn-rich limit set by the measured formation enthalpy. In $\alpha$-MnSe the Mn vacancy is lowest over
most of the window (0.92~eV under Se-rich conditions), a candidate acceptor for its p-type conduction. In
w-MnSe the Se antisite is lowest under Se-rich conditions, narrowly in the A-type order and, by an estimate,
more clearly in the stripe order, and the Se vacancy under Mn-rich conditions. Changing the order shifts the
formation energies by up to 0.26~eV, most for the Mn vacancy. Only the Se vacancy has nearly the same
formation energy in both polymorphs; the other defects differ by 0.4--1.1~eV, although DFT+$U$ places
wurtzite below rock salt, contrary to experiment. Three of the four defects leave an uncompensated moment
of 3--5~$\mu_{\mathrm{B}}$. In the A-type order, band unfolding shows that the sign-reversing splitting of
the most strongly split valence band, about 230~meV and 3~eV below the valence-band maximum (band-edge
states are not resolved), changes its peak height by less than 8\%, the resolution of the analysis, at 3\%
defects, while elsewhere the curve deviates by 18--45~meV (rms).
\end{abstract}


\end{frontmatter}

\section{Introduction}
\label{sec:intro}
Manganese selenide crystallizes in several polymorphs that compete closely in energy. The
stable bulk phase is rock-salt $\alpha$-MnSe, a semiconductor with type-II antiferromagnetic
order~\cite{schlesingerMnSeManganeseseleniumSystem1998,efremdsaLowTemperatureMagnetic2004}.
Wurtzite w-MnSe is metastable; it has been obtained as colloidal
nanocrystals~\cite{sinesColloidalSynthesisNonEquilibrium2010} and as epitaxial
films~\cite{grzybowskiWurtziteVsRocksalt2024}. Tian \textit{et al.} recently
grew both polymorphs as ultrathin nanosheets by chemical vapor deposition, selecting one or the
other through the growth temperature, time and carrier-gas flow, and attributed the selection to
the lower interface energy of wurtzite nuclei, which favors them when the MnCl$_2$ precursor is
scarce~\cite{tianGrowthUltrathinNonlayered2026}. Growth
conditions also fix the native defects, which in turn control doping: rock-salt MnSe is p-type,
with holes as the majority carriers over a wide temperature
range~\cite{prasadElectricalTransportProperties1991,zhengThermoelectricPropertiesPtype2019}.
Its magnetization shows a weak ferromagnetic component and exchange bias, which Pradhan
\textit{et al.} ascribe to intrinsic phase inhomogeneity~\cite{pradhanRevisitingMnSeMagnetic2025};
uncompensated moments from native defects are an alternative. Which native defects have the lowest formation energy in each polymorph, and how this depends on the growth conditions, has not been established for either bulk phase.

The question has acquired a second motivation. Altermagnets are collinear magnets whose
moments are fully compensated, as in an antiferromagnet, but whose bands are spin split without
spin--orbit coupling, because the two spin sublattices are related by a rotation rather than by
a translation or inversion~\cite{smejkalConventionalFerromagnetismAntiferromagnetism2022,
smejkalEmergingResearchLandscape2022}. The splitting has been observed directly in
MnTe~\cite{krempaskyAltermagneticLiftingKramers2024}, and w-MnSe was predicted to be an
altermagnet with a critical temperature above room
temperature~\cite{grzybowskiWurtziteVsRocksalt2024,bezzergaHighPerformanceRoom2025}, in an A-type
order of ferromagnetic layers stacked antiferromagnetically along the $c$ axis. That order has since
been challenged. Mavani \textit{et al.} find a spin-degenerate stripe order lower in energy in all
wurtzite MnX, MnSe included, and reach the A-type phase by Cr
doping~\cite{mavaniCompetingAntiferromagneticPhases2026}. Magnetometry on micrometer-thick MnSe/CdSe superlattices has since shown that w-MnSe orders
antiferromagnetically at $T_N = 62$~K, well below the predicted critical temperatures, but not which order it
adopts~\cite{grzybowskiHighVolumeWurtziteMnSe2026}, while neutron diffraction on wurtzite MnS gives a magnetic cell
larger than the chemical one~\cite{corlissMagneticStructuresPolymorphic1956}, which an A-type order does not have.
We confirm the energy ordering at our settings (Section~\ref{sec:methods}), and we therefore treat the
altermagnetic splitting as a property of the A-type order rather than of the ground state of w-MnSe.

Because that splitting rests on a symmetry relating the two sublattices, it is natural to ask how
robust it is to the point defects present in any real crystal. So far, the question has been addressed
mainly with model Hamiltonians, for impurity scattering and vacancy-induced
magnetization textures~\cite{sukhachovImpurityinducedFriedelOscillations2024,
burkardAnisotropicVacancyinducedMagnetization2026}.
First-principles studies have examined how vacancies and dopants change the magnetism of the
debated altermagnet RuO$_2$~\cite{smolyanyukFragilityMagneticOrder2024,
mavaniConditionsAltermagnetismRuO22026} and of doped
FeSb$_2$~\cite{lucenteCompetitionLocalMagnetic2026}. Gong \textit{et al.} have
classified vacancies, divacancies and antisites in two-dimensional altermagnets, such as V$_2$Se$_2$O, by
the magnetic symmetry that survives relaxation, finding that a single Se vacancy preserves the
$d$-wave splitting. They also report formation energies, in a Supplement not yet public, with no
chemical-potential window in the main text~\cite{gongDefectGeometrySelects2026}. In bulk crystals,
Das \textit{et al.} followed the spin splitting of CrSb through ordered, off-stoichiometric vacancy and
interstitial structures~\cite{dasEffectSymmetryBreaking2026}, and Devaraj \textit{et al.} that of
nickeline MnTe through substitutions on the Te site, calling the lower-symmetry configurations
quasi-altermagnetic~\cite{devarajUnlockingDopingEffects2026}; neither reports formation energies.
Native-defect formation energies do exist for
bulk crystals now classed as altermagnets, for example nickeline MnTe, compared with wurtzite MnTe at the
Mn-rich limit~\cite{hanTemplatedGrowthMetastable2020}, and
hematite~\cite{leeThermodynamicsNativePoint2013}, but those studies do not address the spin splitting, and none of them treats MnSe.

For MnSe, to the best of our knowledge, the only first-principles defect energies are those of Aapro
\textit{et al.} for a monolayer grown on NbSe$_2$, whose structure is neither rock salt nor wurtzite
and in which the defects were evaluated at the Se-rich limit to identify features seen in scanning
tunneling microscopy~\cite{aaproSynthesisPropertiesMonolayer2021}. Related work on 1T-MnSe$_2$ treats
the Se vacancy without formation energies~\cite{erenDefectTolerantDimension2019}. We note that the
term ``Se vacancy'' is also used widely in the electrochemistry literature on MnSe, where it
denotes a synthetic modification of nanostructured electrodes rather than an equilibrium
point defect~\cite{zhong_vacancy_2025}.

Here we compute the four neutral native point defects, the two vacancies and the two
antisites, in bulk $\alpha$-MnSe and w-MnSe with DFT+$U$, using a Hubbard $U$ obtained from
linear response and treating both polymorphs on the same footing. The formation energies span
the full chemical-potential window. We test how the magnetic order of w-MnSe affects them, and we follow
the spin splitting of its A-type order through the defect cells by band unfolding.

\section{Computational details}
\label{sec:methods}

All calculations used spin-polarized density functional theory as implemented in
\textsc{Quantum ESPRESSO}~7.5~\cite{giannozziQUANTUMESPRESSOModular2009,
giannozziAdvancedCapabilitiesMaterials2017a,giannozziUantumESPRESSOExascale2020}, the PBE
exchange--correlation functional~\cite{perdewGeneralizedGradientApproximation1996b}, and
projector-augmented-wave datasets~\cite{blochlProjectorAugmentedwaveMethod1994} from
pslibrary~\cite{dalcorsoPseudopotentialsPeriodicTable2014a} (Mn.pbe-spn-kjpaw\_psl.0.3.1, with $3s3p3d4s$
valence, and Se.pbe-dn-kjpaw\_psl.1.0.0, with $3d4s4p$ valence). Plane-wave cutoffs of 120~Ry (wavefunctions) and 960~Ry (density) converge
the total energy to 0.8~meV/atom relative to 140~Ry in both polymorphs. Brillouin-zone integrals
used $\Gamma$-centered grids~\cite{monkhorstSpecialPointsBrillouinzone1976}:
$8\times8\times8$ for the four-atom rhombohedral cell of $\alpha$-MnSe, $10\times10\times6$
for the four-atom cell of w-MnSe and $3\times3\times3$ for the supercells, with Gaussian
smearing of 0.01~Ry (Supplementary Tables~S1 and S2). Spin--orbit coupling is omitted throughout, since
the altermagnetic spin splitting is non-relativistic~\cite{smejkalConventionalFerromagnetismAntiferromagnetism2022,
smejkalEmergingResearchLandscape2022}.

\subsection{Hubbard correction}
Mn $3d$ states were evaluated using the DFT+$U$ formulation of Dudarev
\textit{et al.}~\cite{dudarevElectronenergylossSpectraStructural1998} and orthogonalized atomic
Hubbard projectors~\cite{timrovPulayForcesDensityfunctional2020}. Published values of $U$ for
Mn in this family span 2.3--6~eV~\cite{aaproSynthesisPropertiesMonolayer2021,
bezzergaHighPerformanceRoom2025,grzybowskiWurtziteVsRocksalt2024}, so $U$ was obtained from
density-functional perturbation theory~\cite{cococcioniLinearResponseApproach2005,
timrovHubbardParametersDensityfunctional2018,timrovHPCodeCalculation2022}. A single
linear-response step on a DFT+$U$ ground state with $U = 5.1$~eV, the value of
Ref.~\cite{bezzergaHighPerformanceRoom2025}, gives $U = 3.76$~eV in A-type w-MnSe, at the lattice of
that reference, and 3.63~eV in $\alpha$-MnSe (eight-atom AFM-I cell; Supplementary Table~S5), identical on
every Mn site of each cell. We did not iterate to self-consistency,
and we use 3.76~eV for both polymorphs and all supercells. A scheme with separate $U$ and $J$ was not
used because \textsc{Quantum ESPRESSO} provides no forces for it, and relaxation around the
defects is essential here. Supplementary Table~S5 gives the sensitivity of the bulk
results for $U = 0$--5~eV; defect formation energies were computed at $U = 3.76$~eV only. The rankings
discussed below have margins of 0.6--1.6~eV, except at the Se-rich end of w-MnSe and near the crossovers.

\subsection{Magnetic order and bulk structures}
\begin{figure*}[t]
\centering
\includegraphics[width=\linewidth]{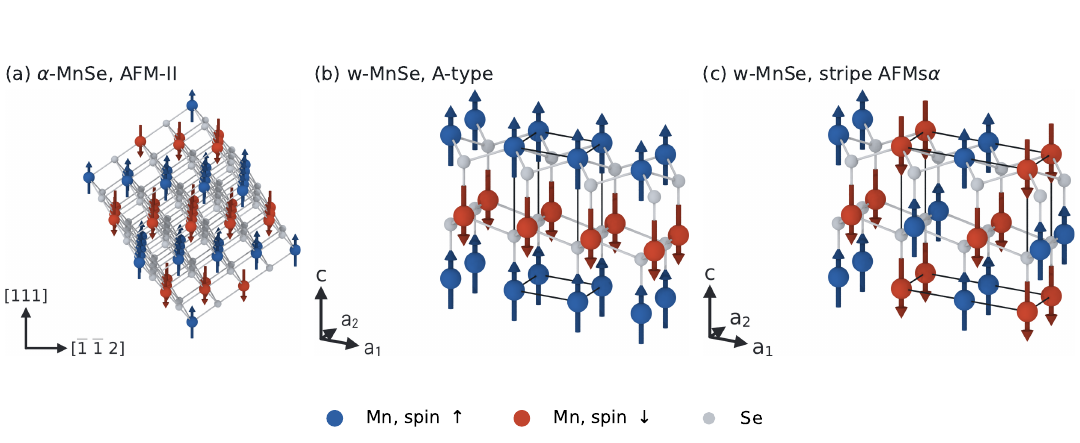}
\caption{Magnetic orders of MnSe. (a) AFM-II order of $\alpha$-MnSe, with ferromagnetic $\{111\}$ planes stacked
antiferromagnetically along [111], which points up the page; $[\bar{1}\bar{1}2]$ lies across it. (b) A-type and (c)
stripe AFMs$\alpha$ orders of w-MnSe, with the lattice vectors $\mathbf{a}_1$, $\mathbf{a}_2$ and $c$. Dark lines in (b)
and (c) mark the magnetic cells; the stripe order doubles the cell along $\mathbf{a}_1$. The direction of the arrows relative to the
lattice is schematic, since the calculations do not include spin--orbit coupling.}
\label{fig:orders}
\end{figure*}

$\alpha$-MnSe is modeled in the type-II antiferromagnetic order (ferromagnetic $\{111\}$
planes stacked antiferromagnetically). Among the ferromagnetic, AFM-I and AFM-II states, it is the lowest, by 23.6~meV per formula unit (one Mn and one Se) at $U = 3.76$~eV (Supplementary Table~S3), in agreement with the
MnO-type order found by neutron diffraction~\cite{efremdsaLowTemperatureMagnetic2004}.
For w-MnSe we use the A-type order (Figure~\ref{fig:orders}), in which the two Mn sublattices, related by the $6_3$ screw axis,
carry antiparallel moments. This is the order of Refs.~\cite{grzybowskiWurtziteVsRocksalt2024,
bezzergaHighPerformanceRoom2025}, for which the altermagnetic splitting is predicted. It is not the
lowest collinear order. In a cell doubled along $\mathbf{a}_1$ the stripe order AFMs$\alpha$ of
Ref.~\cite{mavaniCompetingAntiferromagneticPhases2026}, with ferromagnetic chains alternating within
each layer and two up and two down spins on every Mn tetrahedron, lies 13.0~meV per formula unit below
A-type at the A-type geometry and 14.3~meV below it after relaxing both orders, and the related AFMs$\beta$
lies 2.5~meV below it (Supplementary Table~S3). Our ordering agrees with the 13--19~meV per formula unit obtained in
Ref.~\cite{mavaniCompetingAntiferromagneticPhases2026} at $U = 3$ and 4~eV and with HSE06. All wurtzite
results use the A-type order, the one that carries a spin splitting; Section~\ref{sec:Ef} tests
how the order affects the formation energies. Lattice vectors and internal coordinates of both bulk
cells were fully relaxed.

\subsection{Defect supercells}
Defects were placed in a 64-atom $2\times2\times2$ supercell of the conventional rock-salt
cell, which is commensurate with the AFM-II order, and a 72-atom $3\times3\times2$ supercell of
w-MnSe, both at the relaxed bulk lattice vectors (minimum image distances of 11.1 and 12.8~\AA).
The four neutral defects \VMn, \VSe, \SeMn\ and \MnSe\ were relaxed with the BFGS algorithm at
70/560~Ry, and total energies were recomputed at 120/960~Ry at the relaxed geometries.
Residual forces are below $10^{-4}$~Ry/bohr except for the two wurtzite antisites, which stall and were
accepted at $5\times10^{-4}$~Ry/bohr (Supplementary Table~S6). The relaxations keep the symmetry of the
starting structure, which retains inversion for \VMn\ and \SeMn\ in $\alpha$-MnSe. Symmetry-lowering
distortions and localization of the \VMn\ holes, which occupy degenerate levels, were therefore not explored;
they could only lower \VMn, which would strengthen its lead in $\alpha$-MnSe but could change the Se-rich order
in w-MnSe. The wurtzite cells
started from the ideal internal parameter $u = 0.375$ and relax to $u = 0.377$--0.379 away from the
defect, so the wurtzite host energy is taken at the relaxed bulk value $u = 0.378$ (Supplementary
Section~S3). Interstitials and defect complexes
were not computed, so ``lowest-energy defect'' means the lowest among these four.

Antisites carry frustrated Mn moments, and an unconstrained self-consistent calculation can
settle in a metastable spin state. For \MnSe\ we therefore scanned the total magnetization $M$
on the unrelaxed geometry and relaxed the defect at the minimum: $M = 5~\mu_{\mathrm{B}}$ in rock salt
and $M = 3~\mu_{\mathrm{B}}$ in wurtzite (Supplementary Table~S4). The wurtzite scan visited only one
orientation of the antisite moment relative to its neighbors, so these energies are upper bounds; they
do not affect the ranking, since \MnSe\ is never the lowest. All other defects were unconstrained. Scans
of $M$ at their relaxed geometries find no lower state (Supplementary Table~S12). The bulk hosts are gapped,
with integer occupations. The \VMn\ and \MnSe\ cells are not, nor, slightly, \SeMn\ in $\alpha$-MnSe:
they have partially occupied states, at the valence-band maximum for \VMn\ and near or above the
conduction-band minimum for \MnSe, which give a non-integer $M$ for \VMn\ and a smearing term $-TS$ of up to
0.14~eV. The usual extrapolation to zero width raises their formation energies by at most 0.07~eV
(Supplementary Table~S13). We quote the energies at finite width, which understates the one close margin
(Section~\ref{sec:Ef}).

\subsection{Formation energies and chemical potentials}
The formation energy of a neutral defect $D$ is~\cite{zhangChemicalPotentialDependence1991,
freysoldtFirstprinciplesCalculationsPoint2014}
\begin{equation}
E_f(D) = E_{\mathrm{tot}}(D) - E_{\mathrm{tot}}(\mathrm{host}) - \sum_i n_i \mu_i ,
\label{eq:Ef}
\end{equation}
where $n_i$ atoms of species $i$ are added ($n_i > 0$) or removed ($n_i < 0$). This is the
formalism we used for CrN~\cite{rojasEnergeticsElectronicStructure2018}, and the one Aapro
\textit{et al.} applied to supported monolayer MnSe at the Se-rich
limit~\cite{aaproSynthesisPropertiesMonolayer2021}. Each polymorph is in equilibrium with
itself, $\mu_{\mathrm{Mn}} + \mu_{\mathrm{Se}} = E_{\mathrm{MnSe}}$, with $E_{\mathrm{MnSe}}$ being
the energy per formula unit of that polymorph's host, so that
$E_f(\dmu) = E_f(0) + (n_{\mathrm{Mn}} - n_{\mathrm{Se}})\,\dmu$ with
$\dmu = \mu_{\mathrm{Se}} - \mu_{\mathrm{Se}}^0$. The Se reference $\mu_{\mathrm{Se}}^0$ is
trigonal Se (space group $P3_121$), relaxed with the same settings and an $8\times8\times6$ grid.

The Mn-rich limit, $\dmu = \Delta H_f(\mathrm{MnSe})$, is taken from experiment rather than from
a computed Mn reference. $\alpha$-Mn is a frustrated, non-collinear metal that is paramagnetic
at the temperature where $\Delta H_f$ is measured~\cite{hobbsUnderstandingComplexMetallic2003};
a Hubbard correction is inappropriate for it, and the elemental references we computed place
$\Delta H_f$ anywhere between $-0.17$ and $-2.78$~eV (Supplementary Table~S7). We therefore use
$\Delta H_f = -1.70$~eV, the vapor-decomposition value within the range of $-1.33$ to $-1.85$~eV
compiled by Schlesinger~\cite{schlesingerMnSeManganeseseleniumSystem1998}. Because $E_f$ is linear in
$\dmu$, this choice establishes only the width of the window. No experimental $\Delta H_f$ exists for w-MnSe,
so the same window is used for both polymorphs, which is an assumption (Section~\ref{sec:bulk}). At the
Se-rich end, MnSe + Se $\rightarrow$ MnSe$_2$ is exothermic ($-0.158$~eV from the emf measurements at
427~$^\circ$C compiled in Ref.~\cite{schlesingerMnSeManganeseseleniumSystem1998};
$-0.050$~eV in our calculation for pyrite MnSe$_2$~\cite{hastingsAntiferromagneticStructuresMnS21959},
Supplementary Table~S8), so the accessible Se-rich limit is $\dmu = -0.158$~eV. We also report
$\dmu = 0$, for direct comparison with Ref.~\cite{aaproSynthesisPropertiesMonolayer2021}. Both bounds are
thus measured enthalpies placed relative to the computed $\mu_{\mathrm{Se}}^0$, in the spirit of fitted
elemental reference energies~\cite{lanySemiconductorThermochemistryDensity2008,
stevanovicCorrectingDensityFunctional2012}.

Only neutral defects are considered; charge states and their finite-size
corrections~\cite{freysoldtFirstprinciplesCalculationsPoint2014} are left for future work.
From the unrelaxed \VMn\ of $\alpha$-MnSe, whose formation energy rises by 0.07~eV (0.10~eV at zero
width) from the 64-atom cell to a 128-atom cell (minimum image 15.1~\AA, at 0.6 of the $k$-point density;
Supplementary Section~S3), and
because relaxation (0.28~eV) is short ranged (Section~\ref{sec:local}), we considered $\sim$0.1~eV as the
finite-size uncertainty; no correction is applied. We have no such test for the wurtzite cells.

\subsection{Spin splitting in supercells}
The spin splitting of defective cells was obtained by band
unfolding~\cite{popescuExtractingEffectiveBand2012,kuUnfoldingFirstPrinciplesBand2010}. The path
L$_1$--$\Gamma$--L$_2$, with L$_1 = (\tfrac12,0,\tfrac12)$ and L$_2 = (0,\tfrac12,\tfrac12)$, was sampled
at 41 points of the primitive cell and mapped onto the $3\times3\times2$ supercell so that every
supercell wave vector $\mathbf{K}$ coincides with a primitive $\mathbf{k}$; the spectral weight of
supercell state $n$ at $\mathbf{k} = \mathbf{K}$ is the fraction of its plane-wave norm on the primitive
reciprocal sublattice, evaluated from the pseudo-wavefunction coefficients. For each spin and
wave vector we then follow primitive band 27, the fifth from the top of 31 occupied bands per spin.
Its energy is the weight-averaged energy around the local maximum of the spectral function (Gaussian
width $\sigma = 0.03$~eV) nearest to its pristine energy, with energies aligned through the Se-$4s$ band
(details in Supplementary Section~S5). Considering $s(\mathbf{k}) = E_\uparrow(\mathbf{k}) -
E_\downarrow(\mathbf{k})$ and comparing points at equal distance from $\Gamma$ on the two arms of the
path, the part of $s$ that is odd between the arms is the sign-reversing altermagnetic splitting. For the
pristine supercell this reproduces the primitive-cell splitting at the same $u$, including its
nodes at L$_1$, $\Gamma$ and L$_2$ (Supplementary Table~S10).

\section{Bulk validation}
\label{sec:bulk}

The relaxed AFM-II cell of $\alpha$-MnSe has an equivalent cubic lattice parameter of
5.533~\AA, with a small rhombohedral distortion (pseudo-cubic angle $90.11^\circ$) that
accompanies the AFM-II order. This is 1.3--1.5\% above the measured
5.45--5.46~\AA~\cite{itoMagneticSusceptibilityThermal1978,pradhanRevisitingMnSeMagnetic2025,
efremdsaLowTemperatureMagnetic2004}. For w-MnSe, we obtain $a = 4.271$~\AA,
$c = 6.854$~\AA\ and $u = 0.378$, as compared with $a = 4.178$~\AA\ and $c = 6.783$~\AA\ measured for colloidal wurtzite
MnSe~\cite{sinesColloidalSynthesisNonEquilibrium2010}. Overestimates of this
size are typical of PBE+$U$ and increase with $U$ (Supplementary Table~S5). The absolute
magnetization per Mn, 4.76~$\mu_{\mathrm{B}}$ in $\alpha$-MnSe and 4.92~$\mu_{\mathrm{B}}$ in
w-MnSe, is that of high-spin Mn$^{2+}$.

Both polymorphs are semiconducting at $U = 3.76$~eV, with gaps of 1.42~eV ($\alpha$, AFM-II, indirect)
and 1.69~eV (w, direct); the measured gap of $\alpha$-MnSe is about 2~eV~\cite{zhengThermoelectricPropertiesPtype2019}, and
excitonic transitions in w-MnSe point to a gap near 3.4~eV~\cite{grzybowskiHighVolumeWurtziteMnSe2026}. The gap
difference of about 1.5~eV reported in Ref.~\cite{grzybowskiWurtziteVsRocksalt2024}, whose simple-cubic band
path suggests an AFM-I cell, is recovered (1.46~eV) only with the metastable AFM-I order (gap 0.23~eV); with
AFM-II it is 0.27~eV.

\begin{figure*}[t]
\centering
\includegraphics[width=\linewidth]{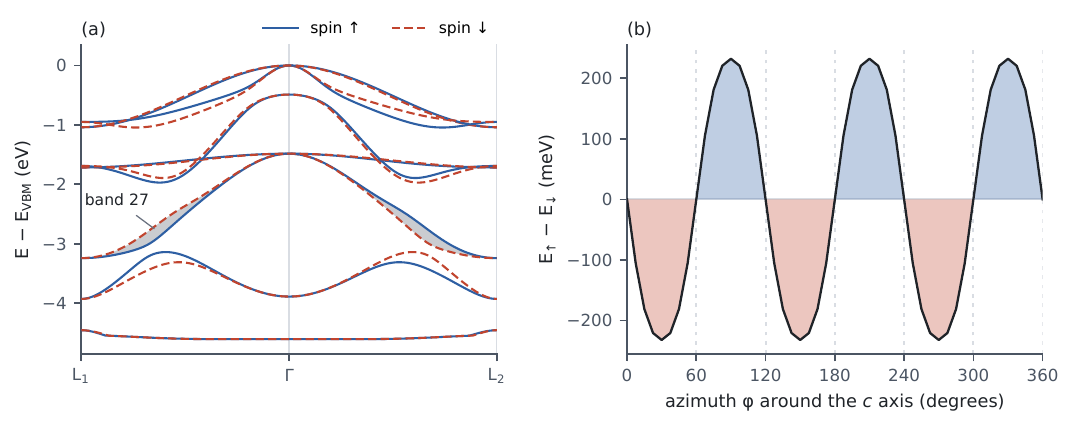}
\caption{Non-relativistic spin splitting of bulk w-MnSe in the A-type order ($U = 3.76$~eV). (a)
Spin-resolved valence bands along L$_1$--$\Gamma$--L$_2$; shading marks the splitting of band 27,
about 3~eV below the valence-band maximum at its peak. (b) Splitting of the same band on a circle through
the peak of (a), at $k_z = 0.198$ and in-plane radius $0.365$ (units of $2\pi/a$): it changes sign every $60^\circ$, the signature of the $6_3$ operation that
exchanges the two Mn sublattices.}
\label{fig:bulksplit}
\end{figure*}

The A-type order of w-MnSe shows the predicted altermagnetic splitting (Figure~\ref{fig:bulksplit}).
Along L$_1$--$\Gamma$--L$_2$ the largest splitting of the occupied bands, 232~meV, lies on band 27, about
3~eV below the valence-band maximum, with opposite sign on the two arms of the path and a sign change every
$60^\circ$ around the $c$ axis (Figure~\ref{fig:bulksplit}b). On the top
valence bands, where Bezzerga \textit{et al.} mark about 245~meV at $U = 5.1$~eV and at their lattice
parameters~\cite{bezzergaHighPerformanceRoom2025}, we find 168~meV at $U = 3.76$~eV. As expected,
$\alpha$-MnSe and the stripe order, whose spin sublattices are related by a translation, show no splitting
(below 0.1~meV).

One bulk quantity is not reproduced. Our calculations place w-MnSe 0.106~eV per formula unit
below $\alpha$-MnSe, whereas experimentally the rock-salt phase is the stable one and wurtzite
MnSe forms only under non-equilibrium or epitaxial conditions~\cite{schlesingerMnSeManganeseseleniumSystem1998,
sinesColloidalSynthesisNonEquilibrium2010,grzybowskiWurtziteVsRocksalt2024}. The stripe order deepens
this by a further 14~meV per formula unit. The inversion does not come from $U$: it is already present in
plain PBE ($-0.066$~eV per formula unit), it deepens only slightly up to $U = 5$~eV ($-0.120$~eV), and
Aapro \textit{et al.} obtained the same sign, about 20~meV per atom, with a different code and
$U$~\cite{aaproSynthesisPropertiesMonolayer2021}, as did Zhou \textit{et al.} for zinc-blende MnSe, which no
$U$ up to 8~eV places above rock salt~\cite{zhouRelativeStabilityElectronic2015}. It may reflect a limitation of the semilocal
description of the competition between octahedral and tetrahedral Mn coordination; a meta-GGA, such as
r$^2$SCAN~\cite{furnessAccurateNumericallyEfficient2020}, or finite-temperature free energies of
the paramagnetic phases, are the natural tests. Each polymorph is referenced to its own host, so a bulk
energy error proportional to the number of Mn atoms cancels in its formation energies. The comparison
between polymorphs is less protected. An error per bond would not cancel, and the shared window is an
assumption: a true polymorph difference $\Delta$ would move the limits of w-MnSe to
$-0.158-\Delta$ and $-1.70+\Delta$.

\section{Formation energies}
\label{sec:Ef}

\input{tab_formation.tex}

\begin{figure*}[t]
\centering
\includegraphics[width=\linewidth]{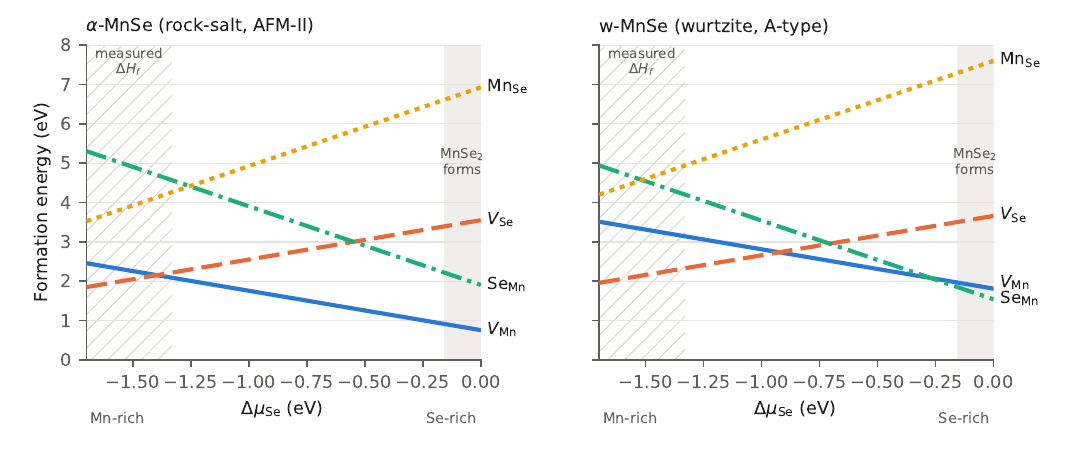}
\caption{Formation energies of the neutral native defects across the chemical-potential window,
from Mn-rich ($\Delta\mu_{\mathrm{Se}} = \Delta H_f = -1.70$~eV) to Se-rich
($\Delta\mu_{\mathrm{Se}} = 0$). In the shaded region MnSe$_2$ is more stable than
MnSe + Se, so the accessible Se-rich limit is $\Delta\mu_{\mathrm{Se}} = -0.158$~eV. The hatched region spans
the measured formation enthalpies down to $-1.33$~eV, where the Mn-rich limit could lie.}
\label{fig:Ef}
\end{figure*}

Table~\ref{tab:Ef} and Figure~\ref{fig:Ef} summarize the formation energies. In $\alpha$-MnSe,
the Mn vacancy is the lowest of these defects over most of the window,
$\Delta\mu_{\mathrm{Se}} > -1.40$~eV. Its formation energy is 0.92~eV at the MnSe$_2$-bounded
Se-rich limit (0.76~eV at the elemental Se limit, $\dmu = 0$) and rises to 2.15~eV at $-1.40$~eV, where the Se
vacancy takes over; the formation energy of \VSe\ is 1.85~eV at the Mn-rich limit. In A-type w-MnSe, the Se antisite and the Mn vacancy are close in energy at the MnSe$_2$ bound (1.86 and 1.97~eV). They are separated by 0.11~eV (0.18~eV at zero smearing width), which is comparable with our uncertainty, and the Se antisite would be lowest only over a narrow part
of the window, $\Delta\mu_{\mathrm{Se}} > -0.27$~eV. The Mn vacancy is lowest between $-0.27$ and
$-0.92$~eV, and the Se vacancy under Mn-rich conditions (1.96~eV at the limit). The Mn antisite is never the lowest: its formation energy is at least 3.5~eV anywhere in the window in either polymorph.

The
lower envelope of the formation energies is highest at $\Delta\mu_{\mathrm{Se}} \approx -1.40$~eV
(2.15~eV) in $\alpha$-MnSe, but at $\Delta\mu_{\mathrm{Se}} \approx -0.92$~eV (2.74~eV) in w-MnSe. These are rankings of neutral
defects. Neutral \VMn\ holds its holes at the valence-band maximum, whereas \VSe\ has occupied states in the gap
(Supplementary Table~S13); in $\alpha$-MnSe no single Fermi level keeps both neutral, so a ranking at a
common Fermi level needs charged states.

The last column of Table~\ref{tab:Ef} compares the polymorphs at fixed $\Delta\mu_{\mathrm{Se}}$. It is
smallest for the Se vacancy ($+0.11$~eV, comparable with the 0.1~eV finite-size uncertainty), the one defect that leaves
the number of Mn atoms unchanged. The other three differ by 0.36--1.05~eV: \VMn\ by
$+1.05$~eV, \MnSe\ by $+0.68$~eV and \SeMn\ by $-0.36$~eV (Supplementary Section~S4 gives them for common
reservoirs). This pattern belongs to the relaxed structures. At ideal sites every defect costs 0.5--1.05~eV
more in wurtzite; the much larger wurtzite relaxations of \VSe\ (0.53~eV, against 0.13~eV in $\alpha$-MnSe)
and \SeMn\ (1.30 against 0.15~eV) reduce their differences, whereas relaxation changes that of \VMn\ by only
0.14~eV (Supplementary Section~S3). Since \VMn\ and \SeMn\ both remove one Mn but differ in sign, we do not
attribute the pattern to the Mn count. Han \textit{et al.} reported the Mn
vacancy to be about 1~eV higher in wurtzite than in nickeline MnTe under Mn-rich
conditions~\cite{hanTemplatedGrowthMetastable2020}. The supported monolayer of
Ref.~\cite{aaproSynthesisPropertiesMonolayer2021} has a Se vacancy at 3.42~eV, close to our 3.55 and 3.66~eV, and a Mn vacancy at 1.37~eV,
between our 0.76 and 1.81~eV. The comparison is qualitative, since that calculation used a different code,
a smaller $U$ (2.3~eV) and a substrate. Aapro \textit{et al.} also found much lower energies for a Mn
interstitial between the monolayer and the substrate (0.16~eV) and for a displaced Mn with a locally
wurtzite-like motif (0.27~eV), which is unstable without the substrate. Neither energy transfers to the bulk.

Table~\ref{tab:Ef} refers to the A-type order of w-MnSe, so we tested how the magnetic order changes it. The
72-atom cell cannot carry the stripe order, so we compared the two orders in a 64-atom $4\times2\times2$ cell at
the A-type lattice (minimum image 8.5~\AA), with each order referenced to its own host (Supplementary Table~S11);
in the A-type order this cell reproduces the 72-atom formation energies of \VMn\ and \SeMn\ to 0.001~eV. With all
atoms relaxed in each order, the stripe order raises \VMn\ by 0.26~eV and leaves \SeMn\ unchanged ($-0.01$~eV).
Without relaxation \VSe\ is also unchanged ($+0.01$~eV); we did not test \MnSe. The stripe \VMn\ holds both holes
in one spin channel ($M = -3.06~\mu_{\mathrm{B}}$, against $-5.27~\mu_{\mathrm{B}}$ in the A-type order); the other
two hole-spin configurations, $M = -5$ and $-7~\mu_{\mathrm{B}}$, lie 0.21 and 0.45~eV higher at its relaxed geometry
(Supplementary Table~S12), more than its whole relaxation energy (0.11~eV). The shift is therefore not a spin-state
artifact; it may be a band-edge effect, which we have not shown (Supplementary Section~S2). If these shifts are added to the values of
Table~\ref{tab:Ef} as estimates, the sequence of lowest-energy defects stays the same but its boundaries move. In
the stripe order \SeMn\ is clearly lowest at the MnSe$_2$ bound (1.85~eV, against 2.23~eV for \VMn), \VMn\ is lowest
only between $-0.54$ and $-0.80$~eV, and \VSe\ remains lowest under Mn-rich conditions (1.97~eV). The polymorph
difference of \VMn\ grows to about $+1.3$~eV. The wurtzite energies of \VMn\ therefore carry an uncertainty of about
0.25~eV from the magnetic order, larger than the finite-size uncertainty; the $\alpha$-MnSe results are not affected.

The low formation energy of \VMn\ in $\alpha$-MnSe connects to transport experiments. Zheng \textit{et al.} associate
the increase in hole concentration on Na doping of p-type MnSe with Mn
vacancies~\cite{zhengThermoelectricPropertiesPtype2019}. Removing a Mn$^{2+}$ ion makes \VMn\ a nominal
double acceptor, so its low energy under all but the most Mn-rich conditions makes it a candidate source
of the p-type conduction; its charged states were not computed.

The main uncertainty in the absolute values is the width of the window. Across the
experimental range of $\Delta H_f$, the Mn-rich vacancy energies move by up to 0.37~eV, while the antisite energies move by up to 0.74~eV, and our computed MnSe$_2$ bound shifts the Se-rich values by at most 0.22~eV
(Supplementary Table~S9). The Mn-rich dominance of \VSe\ in w-MnSe and of \VMn\ at the Se-rich end of
$\alpha$-MnSe are unchanged. In $\alpha$-MnSe, however, the \VMn/\VSe\ crossover at $-1.40$~eV lies inside
the experimental range: for the least negative measured $\Delta H_f$ of $-1.33$~eV, \VMn\ would remain
the lowest-energy defect up to the Mn-rich limit.

\section{Local structure and magnetic moments}
\label{sec:local}

\input{tab_local.tex}

Table~\ref{tab:local} summarizes the relaxations. They are small in rock salt (at most 0.15~\AA) and
larger in wurtzite (up to 0.77~\AA). The Se vacancy draws its Mn neighbors inward, most strongly the Mn on
the $c$ axis of the wurtzite site (0.59~\AA); the other defects push their first shell outward, except
\VMn\ in wurtzite.

The Se antisite shows the clearest structural difference between the polymorphs. In rock salt it stays
at the center of its octahedron, also from an off-center start (Supplementary Section~S3), and pushes its
six Se neighbors out, from 2.77 to 2.86~\AA. In
wurtzite, it leaves the center of its Se$_4$ tetrahedron, moving 0.77~\AA\ along the $c$ axis away from the
apical Se and towards the plane of the three basal Se atoms, which it keeps at 2.60~\AA, while the apical
Se recedes to 3.55~\AA; the site becomes threefold coordinated. This is consistent with \SeMn\ being the one defect with a lower formation energy in w-MnSe than in $\alpha$-MnSe.

The magnetic perturbation is weak and short ranged in all eight cases. Relative to the median moment of
its type in the same cell, no Mn moment beyond the nearest Mn shell changes by more than
0.06~$\mu_{\mathrm{B}}$, apart from a finite-size effect of up to 0.11~$\mu_{\mathrm{B}}$ halfway between a
defect and its periodic image (Supplementary Section~S3). The largest changes, about
0.3~$\mu_{\mathrm{B}}$, are on the frustrated antisite Mn of $\alpha$-MnSe and on the three like-spin
neighbors of the antisite Mn in w-MnSe.

The net magnetization of Table~\ref{tab:local} separates the defects into two classes. \VSe\
leaves the two Mn sublattices intact and the cell compensated ($M = 0$). \VMn\ and \SeMn\ remove
one Mn moment and \MnSe\ adds one, so each leaves the cell with a net moment of
3.0--5.3~$\mu_{\mathrm{B}}$: these defects make both antiferromagnets locally ferrimagnetic. In
$\alpha$-MnSe, where \VMn\ is lowest over most of the window, such uncompensated moments are a candidate source of the weak
ferromagnetic component and the exchange bias reported for nominally antiferromagnetic
MnSe~\cite{pradhanRevisitingMnSeMagnetic2025}. In the A-type order of w-MnSe they break the
compensation that defines the altermagnetic state, which is the question of the next section.

\section{Spin splitting of defective w-MnSe in the A-type order}
\label{sec:splitting}

\input{tab_stage5.tex}

\begin{figure*}[t]
\centering
\includegraphics[width=\linewidth]{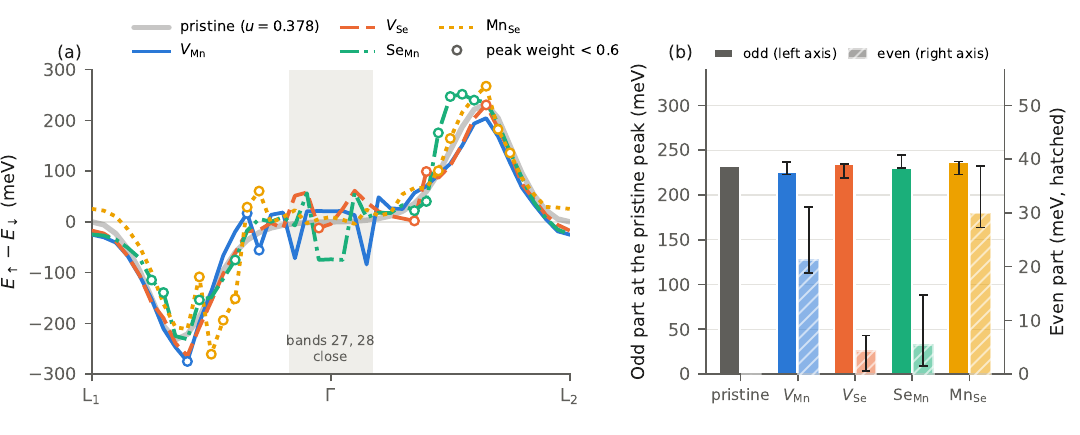}
\caption{Unfolded spin splitting of primitive band 27 of A-type w-MnSe in the 72-atom cell with one
defect. (a) $E_\uparrow - E_\downarrow$ along L$_1$--$\Gamma$--L$_2$, compared with the pristine
curve at the relaxed $u$. Hollow markers show points where the tracked peak holds a weight below 0.6 in at least one spin,
because the band has fragmented. Bands 27 and 28 approach each other in the shaded region, where the values are not
interpreted. (b) Odd (left axis) and even (right axis, hatched) parts at the point where the pristine splitting peaks,
at a Gaussian width of 0.03~eV; error bars show the spread over widths of 0.02--0.05~eV.}
\label{fig:stage5}
\end{figure*}

Table~\ref{tab:stage5} and Figure~\ref{fig:stage5} give the spin splitting of primitive band 27
along L$_1$--$\Gamma$--L$_2$ in the defect cells, all in the A-type order. Each cell contains one defect
per 36 formula units, about 3\% of the sublattice concerned. This is many orders of magnitude above the
equilibrium concentrations implied by the formation energies of Section~\ref{sec:Ef}, so the
calculation is a severe test. The reference is the pristine curve at the relaxed $u = 0.378$ (232~meV),
that of the far field of the relaxed defect cells; the unfolded pristine supercell, at $u = 0.375$,
validates the method and gives 239~meV.

The sign-reversing splitting survives in all the cells. At the point of the path where the pristine
splitting peaks, its odd part is 219--245~meV across the four defects and the Gaussian widths, against
232~meV. This $\pm6$\% is the scatter of the analysis with the width, up to 8\% if the tracking also starts
from the relaxed reference, and we do not resolve a change or a difference between the defects. (For \VSe\
and \MnSe\ the L$_2$-arm point lies just below our 0.6 weight threshold at 0.03~eV, and above it at
0.04--0.05~eV.) Band 26 gives consistent values (Supplementary Section~S5). Off the peak the changes are
larger. The curves differ from the pristine one by 18--45~meV (rms) and by up to 43--120~meV at individual
points where the band fragments (hollow markers in Figure~\ref{fig:stage5}). The peak height hardly depends
on how the band is read off the spectrum (222--243~meV with other estimators, one outlier of 272~meV for
\MnSe; Supplementary Table~S14); the rest of the curve does, and part of the antisite spikes survives every
unbiased estimator, so we do not resolve those changes for the antisites. The defect also makes the two arms
inequivalent, so a measurement on a single arm would see a larger change, up to about 15\%
(Table~\ref{tab:stage5}; 20\% at a width of 0.02~eV). The band remains largely coherent (median weight
0.69--0.87). It fragments, with peak weights down to 0.42--0.44, at 10--11 of the 34 points outside the
region around $\Gamma$ for the antisites, against 3 for the vacancies, consistent with the stronger local
perturbation of the antisites (Section~\ref{sec:local}).

Outside the region around $\Gamma$, where bands 27 and 28 approach each other and the tracked peak merges
them, the defects add a part of $s$ that is even between the arms: the splitting no longer vanishes at L$_1$
and L$_2$, where it is 12--28~meV, and the even part at the peak is at most about 40~meV. This is expected.
The two Mn sublattices of wurtzite lie at different heights along the polar axis, so every operation that
exchanges them includes a translation by $c/2$, and an isolated point defect cannot be invariant under it.
Strictly speaking, none of these cells are altermagnets, including the compensated \VSe\ cell, and an even
splitting is allowed by symmetry; the cells are quasi-altermagnetic in the sense of
Ref.~\cite{devarajUnlockingDopingEffects2026}. This differs from nickeline MnTe, where a single substitution
on the Te site keeps the sublattice-exchanging mirror and roto-inversion~\cite{devarajUnlockingDopingEffects2026},
and from the two-dimensional hosts of Ref.~\cite{gongDefectGeometrySelects2026}, where suitably placed
defects preserve the operation that relates the magnetic sublattices. The odd part persists because most of the crystal is unperturbed and
the Mn moments are only weakly and locally disturbed (Section~\ref{sec:local}), and the part allowed by the
broken symmetry remains much smaller.

Five limitations apply to this statement. The A-type order is not the calculated ground state
(Section~\ref{sec:methods}), so the result concerns that order, reached in
Ref.~\cite{mavaniCompetingAntiferromagneticPhases2026} by Cr doping; the stripe order has no splitting to
preserve. Band 27 lies about 3~eV below the valence-band maximum, in the part of the spectrum where the
splitting is largest; near the band edge the closely spaced bands 30 and 31 merge at our resolution, so we
cannot make the same statement for the states that carry holes. The defects form an ordered lattice at one
concentration rather than a random distribution. They are neutral. And spin--orbit coupling is omitted.

\section{Conclusions}
\label{sec:conclusions}
For the neutral native vacancies and antisites of both polymorphs, three conclusions stand independently of
how w-MnSe is classified. First, the lowest-energy defect depends on both the growth conditions and the
polymorph: Mn vacancies are lowest in $\alpha$-MnSe over most of the window, a candidate source of its p-type
conduction,
whereas in w-MnSe Se antisites are lowest or close to Mn vacancies at the Se-rich end and Se vacancies are lowest under Mn-rich conditions. Second, only the Se vacancy has nearly the same formation energy in both
polymorphs, through its much larger relaxation in wurtzite; the other three defects differ by 0.36--1.05~eV. Third, the perturbation of the magnetic order is weak and
short ranged, but the defects that change the Mn count leave an uncompensated moment of
3--5~$\mu_{\mathrm{B}}$, a candidate source of the weak ferromagnetism reported for nominally
antiferromagnetic MnSe. In w-MnSe, a stripe order lies 14~meV per formula unit below the
A-type order, in line with Ref.~\cite{mavaniCompetingAntiferromagneticPhases2026}; between these orders the formation
energies shift by up to 0.26~eV, most for the Mn vacancy, without changing the sequence of lowest-energy defects.

In the A-type order of w-MnSe, band unfolding showed that the sign-reversing splitting of the most strongly
split valence band, about 230~meV, keeps its peak height to within the 8\% resolution of the analysis at
about 3\% defects, far above equilibrium. This holds true even though no isolated point defect preserves the symmetry behind
it: the defects add an even splitting of a few tens of meV and reduce the coherence of the bands, most
strongly for the antisites, and the rest of the curve deviates by 18--45~meV (rms). The statement
applies to bands about 3~eV below the valence-band maximum; whether the band-edge states behave in the same
way remains open. Charged defects, symmetry-lowering
relaxations, interstitials and the inversion of the polymorph energies also remain open.

\section*{CRediT authorship contribution statement}
\textbf{Tom\'as Rojas Solorzano:} Conceptualization, Methodology, Software, Validation, Investigation,
Formal analysis, Data curation, Visualization, Writing -- original draft, Writing -- review \& editing. \textbf{Dale Igram:} Formal analysis, Writing -- review \& editing.

\section*{Declaration of competing interest}
The authors declare that they have no known competing financial interests or personal relationships that
could have appeared to influence the work reported in this paper.

\section*{Data availability}
The calculation inputs and outputs, the formation-energy and spin-splitting data, the relaxed structures and
the analysis scripts that support this study are openly available in a public
repository~\cite{rojasDataNativeVacancies2026}, the data under CC BY 4.0 and the scripts under the MIT License.

\section*{Acknowledgements}
We gratefully acknowledge the Universidad de Costa Rica High Performance Computing cluster team
for providing computational infrastructure and computing time. This research did not receive any
specific grant from funding agencies in the public, commercial, or not-for-profit sectors.

\section*{Declaration of generative AI and AI-assisted technologies in the manuscript preparation process}
During the preparation of this work the authors used Claude (Anthropic) to draft and revise the
text and to write the scripts that generate the calculation inputs and analyze and plot the results.
No image was generated or altered by an AI tool; the figures were plotted or rendered from the
calculated data by these scripts. After using this tool, the authors reviewed and
edited the content as needed and take full responsibility for the content of the published article.

\bibliographystyle{elsarticle-num}
\bibliography{refs,refs_extra}

\clearpage
\appendix
\section*{Supplementary Material}
\renewcommand{\thetable}{S\arabic{table}}\setcounter{table}{0}

\subsection*{S1. Numerical convergence}
Tables~\ref{tab:S1} and \ref{tab:S2} give the convergence of the bulk total energy with the
plane-wave cutoff and the $k$-point grid. They were run before $U$ was fixed by linear response,
at $U = 5.1$~eV; total-energy convergence with these parameters depends only weakly on $U$. The total
energies of the gapped bulk hosts agree with fixed-occupation calculations to all printed digits.

\subsection*{S2. Magnetic order and the Hubbard $U$}
Table~\ref{tab:S3} compares the magnetic orders at the production $U$ and at $U = 5.1$~eV, including the
stripe orders of w-MnSe in a cell doubled along $\mathbf{a}_1$. Table~\ref{tab:S5} lists the bulk lattice
parameters, magnetization and polymorph energy difference from $U = 0$ to 5~eV, together with the
linear-response values of $U$. The single linear-response step used a $2\times2\times2$ $q$-point mesh
($\chi$ converged to $10^{-5}$), at the lattice of Ref.~\cite{bezzergaHighPerformanceRoom2025}
($a = 4.19$~\AA, $c = 6.68$~\AA, $u = 0.375$) for w-MnSe. Table~\ref{tab:S11} tests how the magnetic order of the wurtzite host
affects the formation energies; in the same 64-atom cell the two hosts differ by 13.0~meV per formula unit.
The cell has a minimum image of 8.54~\AA. In the A-type order it reproduces the relaxed formation energies of
$V_{\mathrm{Mn}}$ and $\mathrm{Se}_{\mathrm{Mn}}$ in the 72-atom cell to 0.001~eV, once the 70$\to$120~Ry correction of
the same defect in $\alpha$-MnSe is added. The stripe $V_{\mathrm{Mn}}$ converges to $M = -3.06~\mu_{\mathrm{B}}$, both
holes in one spin channel; the two other hole configurations lie higher (Table~\ref{tab:S12}). Why the order shifts
$V_{\mathrm{Mn}}$ alone is not established. A band-edge effect is possible: on the $2\times4\times3$ grid the stripe
host has a gap of 2.14~eV, against 1.69~eV for A-type, and its valence-band maximum lies 0.37~eV lower relative to
the Se-$4s$ band; neutral $V_{\mathrm{Mn}}$ holds two holes near that maximum, whereas $V_{\mathrm{Se}}$ holds none.

\subsection*{S3. Defect cells}
The AFM-II order stacks ferromagnetic $\{111\}$ planes, so a cubic $n\times n\times n$
supercell of the conventional rock-salt cell is commensurate with it only for even $n$: the
64-atom $2\times2\times2$ cell is valid, a 216-atom $3\times3\times3$ cell is not, and the next
valid cubic cell has 512 atoms. Larger cells must instead be built as supercells of the AFM-II
primitive cell; the most compact one with a substantially larger defect separation has 128 atoms
(minimum image distance 15.1~\AA, against 11.1~\AA\ for the 64-atom cell). In that cell
(HNF matrix $[[4,0,0],[0,4,0],[0,1,2]]$, $2\times2\times2$ grid, 0.6 of the $k$-point density of the
64-atom $3\times3\times3$ grid, 70/560~Ry) the host energy per
formula unit agrees with the 64-atom cell to 0.28~meV, and the formation energy of the unrelaxed
$V_{\mathrm{Mn}}$ is 0.07~eV higher (0.10~eV after extrapolating to zero smearing width), with the same
magnetic state ($|M| = 3.8$ against 3.6~$\mu_{\mathrm{B}}$; absolute magnetization 4.77 against
4.78~$\mu_{\mathrm{B}}$ per Mn). No such test exists for w-MnSe. The wurtzite defect cells were built at the
ideal $u = 0.375$; a host at that $u$ instead of the relaxed 0.378 would lower every wurtzite formation
energy by 0.03~eV.

Table~\ref{tab:S4} gives the constrained-magnetization scans used to select the spin state of
the Mn antisite, Table~\ref{tab:S12} the scans of the Mn vacancy and Se antisite at their relaxed
geometries, and Table~\ref{tab:S6} the convergence of all eight relaxations. Relaxation from the ideal sites
lowers the energies of $V_{\mathrm{Mn}}$, $V_{\mathrm{Se}}$, $\mathrm{Se}_{\mathrm{Mn}}$ and $\mathrm{Mn}_{\mathrm{Se}}$
by 0.28, 0.13, 0.15 and 0.18~eV in $\alpha$-MnSe and by 0.14, 0.53, 1.30 and 0.55~eV in w-MnSe (70/560~Ry; the
wurtzite values exclude the 0.03~eV the host gains from $u = 0.375$ to 0.378). At ideal sites the polymorph
differences of Table~1 of the main text would therefore be $+0.91$, $+0.51$, $+0.79$ and $+1.05$~eV. Table~\ref{tab:S13} gives the
Kohn--Sham levels and smearing terms of the cells. Site moments are integrated in the Mn spheres of the
PAW construction (bulk values 4.51~$\mu_{\mathrm{B}}$ in $\alpha$-MnSe and 4.43~$\mu_{\mathrm{B}}$ in
w-MnSe). The sphere radius varies with atom type and relaxation, so deviations in a defect cell are taken
relative to the median moment of the same type in that cell.

Beyond the nearest Mn shell no moment deviates by more than 0.06~$\mu_{\mathrm{B}}$, except on Mn atoms
halfway between a defect and its periodic image, which feel both: at 5.53~\AA\ (half the cubic cell edge) for
$V_{\mathrm{Mn}}$ and $\mathrm{Se}_{\mathrm{Mn}}$ in $\alpha$-MnSe, deviating by 0.07--0.11~$\mu_{\mathrm{B}}$,
and at 6.85~\AA\ (half the $c$ edge) for $\mathrm{Se}_{\mathrm{Mn}}$ in w-MnSe, deviating by
0.03~$\mu_{\mathrm{B}}$. The same shell of unrelaxed $V_{\mathrm{Mn}}$ deviates by 0.043~$\mu_{\mathrm{B}}$ in
the 64-atom cell and by 0.010~$\mu_{\mathrm{B}}$ in the 128-atom cell, which identifies this as a finite-size
effect.

The Mn antisite sits in a frustrated environment. On a Se site of rock salt its six Mn neighbors are split
evenly between the two AFM-II sublattices, so its moment (4.16~$\mu_{\mathrm{B}}$) cannot be compensated and
the cell carries $M = 5~\mu_{\mathrm{B}}$. In wurtzite three of its four Mn neighbors belong to one
sublattice; the antisite moment (4.24~$\mu_{\mathrm{B}}$) aligns with them, their moments drop to
4.11~$\mu_{\mathrm{B}}$, and the cell carries $M = 3~\mu_{\mathrm{B}}$.

The relaxations keep the symmetry of the starting structure, which retains inversion for $V_{\mathrm{Mn}}$ and
$\mathrm{Se}_{\mathrm{Mn}}$ in $\alpha$-MnSe. A separate relaxation of $\mathrm{Se}_{\mathrm{Mn}}$ started 0.40~\AA\ off center
along [111], which breaks inversion, returned to the centered structure in 35 steps (largest displacement 0.008~\AA, energy
difference 0.02~meV, from a start 0.26~eV higher). Only this direction was tested.

\subsection*{S4. Chemical potentials}
Table~\ref{tab:S7} lists the formation enthalpy of $\alpha$-MnSe obtained with different
elemental Mn references, which motivates anchoring the Mn-rich limit to experiment. The value used,
$-1.70$~eV, is the vapor-decomposition determination of Ref.~\cite{schlesingerMnSeManganeseseleniumSystem1998};
the midpoint of the compiled range is $-1.59$~eV. The Se reference is plain PBE without van der Waals
corrections, with a volume of 86.9~\AA$^3$ per three atoms, about 6\% above the measured 81.8~\AA$^3$.
Table~\ref{tab:S8} gives the MnSe$_2$ bound of the Se-rich limit, and Table~\ref{tab:S9} the
formation energies at all four limits considered. At fixed $\Delta\mu_{\mathrm{Se}}$ each polymorph has its
own $\mu_{\mathrm{Mn}}$, so the polymorph differences of Table~1 of the main text contain the DFT+$U$ polymorph energy,
0.106~eV per Mn added or removed. With common reservoirs (the same $\mu_{\mathrm{Mn}}$ and $\mu_{\mathrm{Se}}$ for both
polymorphs) they become $+1.16$, $+0.11$, $-0.26$ and $+0.57$~eV for $V_{\mathrm{Mn}}$, $V_{\mathrm{Se}}$,
$\mathrm{Se}_{\mathrm{Mn}}$ and $\mathrm{Mn}_{\mathrm{Se}}$, with the same pattern.

\subsection*{S5. Spin splitting in supercells}
Table~\ref{tab:S10} validates the unfolding analysis on the pristine supercell, where the answer
is known. The supercell eigenvalues are exactly those of the folded primitive cell; pairing
spin-up and spin-down eigenvalues by band index nevertheless returns only 149~meV, because the
energy-sorted lists interleave folded copies whose splittings have opposite signs, and sorted
matching minimizes the largest difference over all pairings. The unfolded spectral weight
recovers the primitive-cell value exactly.

In the defect cells the spectral weight is no longer 0 or 1, and we follow the band as the
local maximum of the spectral function, broadened with a Gaussian of width $\sigma$ (0.03~eV unless
stated), that lies nearest to its pristine energy, within $\pm0.20$~eV and with a height of at least 0.35
of a coherent band. Its energy is the weight-averaged energy within $\pm2.5\sigma$ of that maximum, and the
energies of each cell are aligned through its weight-averaged Se-$4s$ band. The result does not change for
search windows between $\pm0.15$ and $\pm0.40$~eV. When summed over the path outside the region around
$\Gamma$, the odd part of the splitting is 0.91--1.04 of its pristine value.

The spikes in Figure~4(a) lie where a band fragments into peaks of similar weight and the tracker follows one of
them. Table~\ref{tab:S14} compares ways of avoiding this. A centroid in a hard window is exact on the pristine cell, but it
changes abruptly as states enter or leave the window. A centroid in a Gaussian window gives the smoothest curves, but it is
pulled towards the pristine energy on which it is centered, by 25--41\% of any offset. If the center is instead iterated to
self-consistency, the bias goes away and the antisite spikes largely return, so they are not only an artifact of tracking. We
therefore keep the tracked peak, mark the points of low weight, and rely only on the peak height, which these estimators agree on.

Band 26 reaches 230~meV in the bulk at the 41-point sampling. Its unfolded values at the peak are 213--231~meV at a width of
0.03~eV, against 230~meV, but in the $V_{\mathrm{Se}}$ and $\mathrm{Se}_{\mathrm{Mn}}$ cells it is more
fragmented and its value depends more on the width. Band 30, which carries 168~meV in the bulk, merges with
band 31 at our resolution and cannot be tracked.

\input{si_tables.tex}

\end{document}

%% file: tab_formation.tex
\begin{table*}[t]
\centering
\caption{Formation energies (eV) of the neutral native defects in $\alpha$-MnSe (64-atom cell)
and w-MnSe (72-atom cell). Se-rich: elemental trigonal Se ($\Delta\mu_{\mathrm{Se}} = 0$) and the
MnSe$_2$-bounded limit ($\Delta\mu_{\mathrm{Se}} = -0.158$~eV); Mn-rich:
$\Delta\mu_{\mathrm{Se}} = \Delta H_f = -1.70$~eV. The slope
$n_{\mathrm{Mn}} - n_{\mathrm{Se}}$ of $E_f$ versus $\Delta\mu_{\mathrm{Se}}$ is listed, and the
last column gives the polymorph difference $E_f(\mathrm{w}) - E_f(\alpha)$, which is the same in
every limit.}
\label{tab:Ef}
\small\setlength{\tabcolsep}{4pt}
\begin{tabular}{lrrrrrrrr}
\toprule
 & & \multicolumn{3}{c}{$\alpha$-MnSe} & \multicolumn{3}{c}{w-MnSe} & \\
\cmidrule(lr){3-5}\cmidrule(lr){6-8}
Defect & $n_{\mathrm{Mn}}-n_{\mathrm{Se}}$ & Se & MnSe$_2$ & Mn-rich & Se & MnSe$_2$ & Mn-rich & w$-\alpha$ \\
\midrule
$V_{\mathrm{Mn}}$ & $-1$ & 0.76 & 0.92 & 2.46 & 1.81 & 1.97 & 3.51 & $+1.05$ \\
$V_{\mathrm{Se}}$ & $+1$ & 3.55 & 3.39 & 1.85 & 3.66 & 3.50 & 1.96 & $+0.11$ \\
$\mathrm{Se}_{\mathrm{Mn}}$ & $-2$ & 1.90 & 2.22 & 5.30 & 1.54 & 1.86 & 4.94 & $-0.36$ \\
$\mathrm{Mn}_{\mathrm{Se}}$ & $+2$ & 6.92 & 6.61 & 3.52 & 7.60 & 7.28 & 4.20 & $+0.68$ \\
\bottomrule
\end{tabular}
\end{table*}

%% file: tab_local.tex
\begin{table}[t]
\centering
\caption{Local relaxation and net magnetization of the defect cells. $\Delta r_1$: mean radial
displacement of the first coordination shell (positive outward); $|u|_{\max}$: largest atomic
displacement in the cell; $|M|$: net magnetization of the cell ($\mu_{\mathrm{B}}$).
First shells: six neighbors at 2.77~\AA\ in $\alpha$-MnSe, four at 2.57--2.61~\AA\ in w-MnSe.
$^a$Constrained at the minimum of the $M$ scan.}
\label{tab:local}
\begin{tabular}{lrrrrrr}
\toprule
 & \multicolumn{3}{c}{$\alpha$-MnSe} & \multicolumn{3}{c}{w-MnSe} \\
\cmidrule(lr){2-4}\cmidrule(lr){5-7}
Defect & $\Delta r_1$ (\AA) & $|u|_{\max}$ (\AA) & $|M|$ & $\Delta r_1$ (\AA) & $|u|_{\max}$ (\AA) & $|M|$ \\
\midrule
$V_{\mathrm{Mn}}$ & $+0.129$ & 0.129 & 3.71 & $-0.027$ & 0.070 & 5.25 \\
$V_{\mathrm{Se}}$ & $-0.148$ & 0.148 & 0.00 & $-0.390$ & 0.592 & 0.00 \\
$\mathrm{Se}_{\mathrm{Mn}}$ & $+0.092$ & 0.092 & 3.00 & $+0.166$ & 0.766 & 5.00 \\
$\mathrm{Mn}_{\mathrm{Se}}$ & $+0.106$ & 0.113 & 5$^a$ & $+0.298$ & 0.317 & 3$^a$ \\
\bottomrule
\end{tabular}
\end{table}

%% file: tab_stage5.tex
\begin{table}[t]
\centering
\caption{Unfolded spin splitting of primitive band 27 of A-type w-MnSe in the 72-atom cell (meV), at the
point of the path where the pristine splitting peaks: part odd between the arms $s_{\mathrm{odd}}$,
the signed splitting on the two arms, and the even part $s_{\mathrm{even}}$; also the
splitting at L$_{1,2}$, the rms deviation from the pristine curve at $u = 0.378$ (at $u = 0.375$ for the
pristine supercell itself), and the median spectral weight $w$ of the tracked peak (1 for a coherent
band), both over the path outside the region around $\Gamma$. Ranges: Gaussian widths 0.02--0.05~eV;
arms, rms and $w$ at 0.03~eV. At 0.02~eV the track is lost at 2 ($\mathrm{Se}_{\mathrm{Mn}}$) and 4
($\mathrm{Mn}_{\mathrm{Se}}$) of 41 points. $|M|$: net moment of the cell ($\mu_{\mathrm{B}}$). $^a$Constrained.}
\label{tab:stage5}
\small\setlength{\tabcolsep}{4pt}
\begin{tabular}{lrrrrrrr}
\toprule
Cell & $|M|$ & $s_{\mathrm{odd}}$ & Arms & $|s_{\mathrm{even}}|$ & $|s(\mathrm{L}_{1,2})|$ & rms & $w$ \\
\midrule
Pristine, $u = 0.378$ & 0 & 232 & $-232$/$+232$ & 0 & 0 & -- & 1.00 \\
Pristine, $u = 0.375$ & 0 & 239 & $-239$/$+239$ & 0 & 0 & 0 & 1.00 \\
$V_{\mathrm{Mn}}$ & 5.25 & 223--237 & $-247$/$+204$ & 19--31 & 18--25 & 27 & 0.87 \\
$V_{\mathrm{Se}}$ & 0.00 & 219--235 & $-239$/$+230$ & 1--7 & 12--19 & 18 & 0.84 \\
$\mathrm{Se}_{\mathrm{Mn}}$ & 5.00 & 230--245 & $-225$/$+236$ & 2--15 & 20--25 & 32 & 0.73 \\
$\mathrm{Mn}_{\mathrm{Se}}$ & 3$^a$ & 223--237 & $-208$/$+267$ & 27--39 & 14--28 & 45 & 0.69 \\
\bottomrule
\end{tabular}
\end{table}

%% file: si_tables.tex
\begin{table}[h]
\centering
\caption{Plane-wave cutoff convergence of the bulk total energy (relative to 140~Ry,
meV/atom) at fixed geometry, with $E_{\mathrm{cut}}^{\rho} = 8E_{\mathrm{cut}}^{\psi}$.
8-atom $\alpha$-MnSe and 4-atom w-MnSe cells, $U = 5.1$~eV (before $U$ was determined).}
\label{tab:S1}
\begin{tabular}{rrr}
\toprule
$E_{\mathrm{cut}}^{\psi}$ (Ry) & $\Delta E$ ($\alpha$) & $\Delta E$ (w) \\
\midrule
70 & $24.48$ & $24.14$ \\
80 & $8.95$ & $8.86$ \\
90 & $5.76$ & $5.65$ \\
100 & $4.24$ & $4.18$ \\
110 & $2.26$ & $2.24$ \\
120 & $0.83$ & $0.80$ \\
130 & $0.21$ & $0.17$ \\
140 & $0.00$ & $0.00$ \\
\bottomrule
\end{tabular}
\par\smallskip\parbox{0.9\linewidth}{\footnotesize Production energies use 120~Ry (0.8~meV/atom).}
\end{table}

\begin{table}[h]
\centering
\caption{$k$-point convergence of the bulk total energy (meV/atom, relative to the densest
grid). Same cells and $U$ as Table~\ref{tab:S1}, 70~Ry. The $3\times3\times3$ grids of the 64- and
72-atom supercells correspond to $6\times6\times6$ ($\alpha$) and $9\times9\times6$ (w) in these cells.}
\label{tab:S2}
\begin{tabular}{llr}
\toprule
Phase & Grid & $\Delta E$ \\
\midrule
$\alpha$ & 4$\times$4$\times$4 & $0.618$ \\
$\alpha$ & 6$\times$6$\times$6 & $0.007$ \\
$\alpha$ & 8$\times$8$\times$8 & $-0.005$ \\
$\alpha$ & 10$\times$10$\times$10 & $0.000$ \\
w & 6$\times$6$\times$4 & $0.364$ \\
w & 8$\times$8$\times$6 & $0.034$ \\
w & 10$\times$10$\times$6 & $0.003$ \\
w & 12$\times$12$\times$8 & $0.000$ \\
\bottomrule
\end{tabular}
\end{table}

\begin{table}[h]
\centering
\caption{Relative energies of magnetic orders (meV per formula unit) and absolute
magnetization per Mn ($\mu_{\mathrm{B}}$). $\alpha$-MnSe: 64-atom cells; w-MnSe: 4-atom cell, and an 8-atom
cell doubled along $\mathbf{a}_1$ ($5\times10\times6$ grid) at the geometry of the relaxed A-type order. In
AFMs$\alpha$ every Mn tetrahedron has two up and two down spins; in AFMs$\beta$ and A-type three and one
\cite{mavaniCompetingAntiferromagneticPhases2026}.}
\label{tab:S3}
\begin{tabular}{lllrr}
\toprule
Phase & $U$ & Order & $\Delta E$ & $|m|$/Mn \\
\midrule
$\alpha$ & $U = 3.76$~eV, 70~Ry & AFM-II & $0.0$ & $4.77$ \\
$\alpha$ & $U = 3.76$~eV, 70~Ry & AFM-I & $23.6$ & $4.90$ \\
$\alpha$ & $U = 3.76$~eV, 70~Ry & FM & $32.7$ & $5.00$ \\
$\alpha$ & $U = 5.1$~eV, 120~Ry & AFM-II & $0.0$ & $4.80$ \\
$\alpha$ & $U = 5.1$~eV, 120~Ry & AFM-I & $15.9$ & $4.92$ \\
$\alpha$ & $U = 5.1$~eV, 120~Ry & FM & $23.2$ & $5.00$ \\
w & $U = 3.76$~eV, 120~Ry & A-type (altermagnetic) & $0.0$ & $4.92$ \\
w & $U = 3.76$~eV, 120~Ry & FM & $31.7$ & $5.00$ \\
w & $U = 5.1$~eV, 120~Ry & A-type (altermagnetic) & $0.0$ & $4.93$ \\
w & $U = 5.1$~eV, 120~Ry & FM & $31.5$ & $5.00$ \\
w & $U = 3.76$~eV, 120~Ry, doubled cell & Stripe AFMs$\alpha$ & $0.0$ & $4.89$ \\
w & $U = 3.76$~eV, 120~Ry, doubled cell & Stripe AFMs$\beta$ & $10.5$ & $4.91$ \\
w & $U = 3.76$~eV, 120~Ry, doubled cell & A-type (altermagnetic) & $13.0$ & $4.92$ \\
w & $U = 3.76$~eV, 120~Ry, doubled cell & FM & $44.8$ & $5.00$ \\
\bottomrule
\end{tabular}
\par\smallskip\parbox{0.9\linewidth}{\footnotesize The A-type energy in the doubled cell reproduces twice the 4-atom value to $2\times10^{-8}$~Ry. With full relaxation of cell and atoms, AFMs$\alpha$ lies 14.3~meV per formula unit below A-type, its cell distorting to $a_1 = 4.235$, $a_2 = 4.260$, $c = 6.906$~\AA, $\gamma = 120.2^\circ$ (A-type $a = 4.271$, $c = 6.855$~\AA).}
\end{table}

\begin{table}[h]
\centering
\caption{Constrained total-magnetization scan for the Mn antisite (meV, relative to the
minimum), on the unrelaxed geometry at 70~Ry. The defects were relaxed at the minima,
$M = 5~\mu_{\mathrm{B}}$ ($\alpha$) and $3~\mu_{\mathrm{B}}$ (w).}
\label{tab:S4}
\begin{tabular}{lrr}
\toprule
Phase & $M$ ($\mu_{\mathrm{B}}$) & $\Delta E$ \\
\midrule
$\alpha$ & $0$ & $2411$ \\
$\alpha$ & $1$ & $1488$ \\
$\alpha$ & $2$ & $620$ \\
$\alpha$ & $3$ & $415$ \\
$\alpha$ & $4$ & $314$ \\
$\alpha$ & $5$ & $0$ \\
$\alpha$ & $6$ & $62$ \\
$\alpha$ & $7$ & $110$ \\
$\alpha$ & $8$ & $704$ \\
$\alpha$ & $9$ & $1332$ \\
w & $2$ & $783$ \\
w & $3$ & $0$ \\
w & $4$ & $101$ \\
w & $5$ & $278$ \\
w & $6$ & $313$ \\
\bottomrule
\end{tabular}
\end{table}

\begin{table}[h]
\centering
\caption{Dependence of the bulk properties on $U$ (both cells fully relaxed at 120~Ry):
equivalent cubic lattice parameter of AFM-II $\alpha$-MnSe, lattice parameters of w-MnSe (\AA),
absolute magnetization per Mn ($\mu_{\mathrm{B}}$), and $\Delta E = E(\mathrm{w}) - E(\alpha)$
(eV per formula unit).}
\label{tab:S5}
\begin{tabular}{rrrrrrr}
\toprule
$U$ (eV) & $a_{\alpha}$ & $a_{\mathrm{w}}$ & $c_{\mathrm{w}}$ & $|m|_{\alpha}$ & $|m|_{\mathrm{w}}$ & $\Delta E$ \\
\midrule
0 & $5.4029$ & $4.1774$ & $6.6433$ & $4.43$ & $4.61$ & $-0.0658$ \\
1.0 & $5.4454$ & $4.2095$ & $6.7182$ & $4.55$ & $4.73$ & $-0.0732$ \\
2.0 & $5.4807$ & $4.2347$ & $6.7751$ & $4.64$ & $4.81$ & $-0.0844$ \\
2.3 & $5.4906$ & $4.2418$ & $6.7912$ & $4.67$ & $4.83$ & $-0.0881$ \\
3.0 & $5.5115$ & $4.2563$ & $6.8228$ & $4.72$ & $4.88$ & $-0.0968$ \\
3.76 & $5.5325$ & $4.2706$ & $6.8543$ & $4.76$ & $4.92$ & $-0.1062$ \\
5.0 & $5.5663$ & $4.2919$ & $6.8992$ & $4.83$ & $4.97$ & $-0.1202$ \\
\bottomrule
\end{tabular}
\par\smallskip\parbox{0.9\linewidth}{\footnotesize Linear response (hp.x), one step from a $U = 5.1$~eV ground state: $U = 3.7574$~eV (w-MnSe, all sites) and $3.6285$~eV ($\alpha$-MnSe, AFM-I 8-atom cell, all sites). Experiment: $a_{\alpha} = 5.452$--5.464~\AA; colloidal w-MnSe $a = 4.178$, $c = 6.783$~\AA.}
\end{table}

\begin{table}[h]
\centering
\caption{Convergence of the defect relaxations (70~Ry): number of ionic steps, largest
residual force component (Ry/bohr), and spread of the total energy over the last four steps
(meV). Accepted either by BFGS convergence at $10^{-4}$~Ry/bohr or by criterion (ii): residual
force $\le 5\times10^{-4}$~Ry/bohr and energy constant to 0.1~meV.}
\label{tab:S6}
\begin{tabular}{llrrrl}
\toprule
Phase & Defect & Steps & $F_{\max}$ & $\delta E_4$ & Accepted by \\
\midrule
$\alpha$ & $V_{\mathrm{Mn}}$ & 24 & $9.8\times10^{-5}$ & $0.039$ & BFGS \\
$\alpha$ & $V_{\mathrm{Se}}$ & 19 & $8.6\times10^{-5}$ & $0.039$ & BFGS \\
$\alpha$ & $\mathrm{Se}_{\mathrm{Mn}}$ & 13 & $5.1\times10^{-5}$ & $0.026$ & BFGS \\
$\alpha$ & $\mathrm{Mn}_{\mathrm{Se}}$ & 38 & $4.8\times10^{-5}$ & $0.013$ & BFGS \\
w & $V_{\mathrm{Mn}}$ & 31 & $9.8\times10^{-5}$ & $0.075$ & BFGS \\
w & $V_{\mathrm{Se}}$ & 29 & $9.1\times10^{-5}$ & $0.014$ & BFGS \\
w & $\mathrm{Se}_{\mathrm{Mn}}$ & 39 & $2.1\times10^{-4}$ & $0.027$ & criterion (ii) \\
w & $\mathrm{Mn}_{\mathrm{Se}}$ & 16 & $3.2\times10^{-4}$ & $0.095$ & criterion (ii) \\
\bottomrule
\end{tabular}
\par\smallskip\parbox{0.9\linewidth}{\footnotesize The w-MnSe Mn$_{\mathrm{Se}}$ relaxation was run at the constrained moment $M = 3~\mu_{\mathrm{B}}$ after the scan of Table~\ref{tab:S4}; steps are counted for that run only. For w-MnSe $V_{\mathrm{Se}}$ and $\mathrm{Se}_{\mathrm{Mn}}$ the output of the first leg of the relaxation is not preserved, so the step counts are lower bounds.}
\end{table}

\begin{table}[h]
\centering
\caption{Formation enthalpy of $\alpha$-MnSe, $\Delta H_f = E_{\mathrm{MnSe}} -
\mu_{\mathrm{Mn}}^0 - \mu_{\mathrm{Se}}^0$ (eV per formula unit), for different elemental Mn
references, with the absolute magnetization per Mn of the reference ($\mu_{\mathrm{B}}$).}
\label{tab:S7}
\begin{tabular}{lrr}
\toprule
Mn reference & $|m|$/Mn & $\Delta H_f$ \\
\midrule
bcc Mn, $U = 3.76$~eV, Gaussian & $4.66$ & $-1.800$ \\
bcc Mn, $U = 0$, Gaussian & $2.42$ & $-0.258$ \\
$\alpha$-Mn (58 at.), non-magnetic, $U = 0$ & $0.00$ & $-0.168$ \\
$\alpha$-Mn (58 at.), collinear AFM, $U = 3.76$~eV & $3.76$ & $-2.782$ \\
$\alpha$-Mn (58 at.), collinear AFM, $U = 0$ & -- & not converged \\
Experiment (determinations compiled in Ref.~\cite{schlesingerMnSeManganeseseleniumSystem1998}) & -- & $-1.33$ to $-1.85$ \\
Value used (Mn-rich limit) &  & $-1.70$ \\
\bottomrule
\end{tabular}
\par\smallskip\parbox{0.9\linewidth}{\footnotesize The only references that come close to experiment carry a moment (4.66~$\mu_{\mathrm{B}}$) far above the 0.5--1.9~$\mu_{\mathrm{B}}$ of $\alpha$-Mn; $\alpha$-Mn with $U$ is 0.98~eV/atom above bcc Mn, the wrong phase ordering. The Mn-rich limit is therefore anchored to experiment (main text, Sec.~2).}
\end{table}

\begin{table}[h]
\centering
\caption{Competing phase at the Se-rich limit: pyrite MnSe$_2$ (collinear AFM-I,
$U = 3.76$~eV, fully relaxed) and the reaction energy of MnSe + Se $\rightarrow$ MnSe$_2$ (eV).}
\label{tab:S8}
\begin{tabular}{lrr}
\toprule
Quantity & This work & Experiment \\
\midrule
Lattice parameter (\AA) & $6.530$ & 6.429 \\
$|m|$/Mn ($\mu_{\mathrm{B}}$) & $4.83$ & -- \\
$\Delta H$(MnSe + Se $\rightarrow$ MnSe$_2$) & $-0.050$ & $-0.158$ \\
\bottomrule
\end{tabular}
\par\smallskip\parbox{0.9\linewidth}{\footnotesize Experimental reaction energy from the emf measurements of both compounds compiled in Ref.~\cite{schlesingerMnSeManganeseseleniumSystem1998}. MnSe$_2$ is helimagnetic below 49~K; the collinear AFM-I order is an approximation.}
\end{table}

\begin{table}[h]
\centering
\caption{Formation energies (eV) at all four chemical-potential limits considered:
elemental Se, the MnSe$_2$ bound from experiment and from our calculation, and the Mn-rich limit.}
\label{tab:S9}
\begin{tabular}{llrrrr}
\toprule
Phase & Defect & Se ($\Delta\mu=0$) & MnSe$_2$ (expt, $-0.158$) & MnSe$_2$ (DFT, $-0.050$) & Mn-rich ($-1.70$) \\
\midrule
$\alpha$ & $V_{\mathrm{Mn}}$ & 0.76 & 0.92 & 0.81 & 2.46 \\
$\alpha$ & $V_{\mathrm{Se}}$ & 3.55 & 3.39 & 3.50 & 1.85 \\
$\alpha$ & $\mathrm{Se}_{\mathrm{Mn}}$ & 1.90 & 2.22 & 2.00 & 5.30 \\
$\alpha$ & $\mathrm{Mn}_{\mathrm{Se}}$ & 6.92 & 6.61 & 6.82 & 3.52 \\
w & $V_{\mathrm{Mn}}$ & 1.81 & 1.97 & 1.86 & 3.51 \\
w & $V_{\mathrm{Se}}$ & 3.66 & 3.50 & 3.61 & 1.96 \\
w & $\mathrm{Se}_{\mathrm{Mn}}$ & 1.54 & 1.86 & 1.64 & 4.94 \\
w & $\mathrm{Mn}_{\mathrm{Se}}$ & 7.60 & 7.28 & 7.50 & 4.20 \\
\bottomrule
\end{tabular}
\end{table}

\begin{table}[h]
\centering
\caption{Validation of the supercell spin-splitting analysis on the pristine 72-atom w-MnSe
cell (peak splitting of band 27 along L$_1$--$\Gamma$--L$_2$, meV). The
supercell is built at the ideal internal parameter, so its reference is the ideal-$u$
primitive value.}
\label{tab:S10}
\begin{tabular}{lr}
\toprule
Analysis & Peak splitting \\
\midrule
Primitive cell, relaxed $u = 0.378$ & 232.3 \\
Primitive cell, ideal $u = 0.375$ & 239.0 \\
72-atom supercell, eigenvalues paired by band index & 149.0 \\
Folded primitive spectrum, paired by band index & 149.1 \\
72-atom supercell, unfolded, peak tracking (this work) & 239.0 \\
\bottomrule
\end{tabular}
\par\smallskip\parbox{0.9\linewidth}{\footnotesize The supercell eigenvalues coincide with the folded primitive spectrum to 0.4~meV, so the deficit of the index-paired value is entirely an artefact of pairing in a folded spectrum. Unfolded: exactly 8 states per spin carry weight $>0.5$ in the window at every $k$, the splitting vanishes at L$_1$, $\Gamma$ and L$_2$, and its uniform (even) part is zero.}
\end{table}

\begin{table}[h]
\centering
\caption{Effect of the magnetic order of w-MnSe on the formation energies, in a 64-atom
$4\times2\times2$ supercell, which carries both orders (the 72-atom production cell has an odd repeat along
$\mathbf{a}_1$ and cannot carry the stripe order), at the relaxed A-type lattice and $u = 0.378$, $2\times4\times3$
grid, 70/560~Ry. $\delta E_f = E_f(\mathrm{AFMs}\alpha) - E_f(\mathrm{A\text{-}type})$ in eV at fixed
$\Delta\mu_{\mathrm{Se}}$, each order referenced to its own host; $M$ in $\mu_{\mathrm{B}}$ for A-type and stripe.
Unrelaxed: ideal sites. Relaxed: all atoms relaxed in each order, including the stripe host.}
\label{tab:S11}
\begin{tabular}{lrrrrrr}
\toprule
& \multicolumn{3}{c}{Unrelaxed} & \multicolumn{3}{c}{Relaxed} \\ Defect & $M_{\mathrm{A}}$ & $M_{\mathrm{s}}$ & $\delta E_f$ & $M_{\mathrm{A}}$ & $M_{\mathrm{s}}$ & $\delta E_f$ \\
\midrule
$V_{\mathrm{Mn}}$ & $-5.39$ & $-3.05$ & $+0.240$ & $-5.27$ & $-3.06$ & $+0.261$ \\
$V_{\mathrm{Se}}$ & $+0.00$ & $+0.00$ & $+0.006$ & -- & -- & -- \\
$\mathrm{Se}_{\mathrm{Mn}}$ & $-5.02$ & $-3.01$ & $-0.053$ & $-5.00$ & $-5.00$ & $-0.008$ \\
\bottomrule
\end{tabular}
\par\smallskip\parbox{0.9\linewidth}{\footnotesize $\mathrm{Mn}_{\mathrm{Se}}$ was not tested, and $V_{\mathrm{Se}}$ was not relaxed. Relaxed residual forces are below $10^{-4}$~Ry/bohr, except stripe $V_{\mathrm{Mn}}$ ($2\times10^{-4}$~Ry/bohr, energy constant to 0.1~meV over its last 10 steps). The relaxed A-type $\mathrm{Se}_{\mathrm{Mn}}$ was run at fixed $M = -5$, the state its unconstrained cycle converges to. Relaxing the stripe host lowers it by a further 0.3~meV per formula unit. Relaxation lowers $V_{\mathrm{Mn}}$ by 0.12~eV in the A-type order and by 0.11~eV in the stripe order. Its stripe spin state is tested in Table~\ref{tab:S12}.}
\end{table}

\begin{table}[h]
\centering
\caption{Constrained-magnetization scans at the relaxed geometry of the unconstrained defect
(70/560~Ry; meV, relative to the unconstrained state).}
\label{tab:S12}
\begin{tabular}{llrr}
\toprule
Phase & Defect & $M$ ($\mu_{\mathrm{B}}$) & $\Delta E$ \\
\midrule
$\alpha$ & $V_{\mathrm{Mn}}$ & $+3.71$ (unconstrained) & $0$ \\
$\alpha$ & $V_{\mathrm{Mn}}$ & $+3$ & $+33$ \\
$\alpha$ & $V_{\mathrm{Mn}}$ & $+5$ & $+68$ \\
$\alpha$ & $\mathrm{Se}_{\mathrm{Mn}}$ & $+3.00$ (unconstrained) & $0$ \\
$\alpha$ & $\mathrm{Se}_{\mathrm{Mn}}$ & $+1$ & $+1074$ \\
$\alpha$ & $\mathrm{Se}_{\mathrm{Mn}}$ & $+5$ & $+266$ \\
$\alpha$ & $\mathrm{Se}_{\mathrm{Mn}}$ & $+7$ & $+382$ \\
w & $V_{\mathrm{Mn}}$ & $-5.26$ (unconstrained) & $0$ \\
w & $V_{\mathrm{Mn}}$ & $-3$ & $+304$ \\
w & $V_{\mathrm{Mn}}$ & $-5$ & $+4$ \\
w & $V_{\mathrm{Mn}}$ & $-7$ & $+204$ \\
w & $\mathrm{Se}_{\mathrm{Mn}}$ & $-5.00$ (unconstrained) & $0$ \\
w & $\mathrm{Se}_{\mathrm{Mn}}$ & $-1$ & $+3131$ \\
w & $\mathrm{Se}_{\mathrm{Mn}}$ & $-3$ & $+1185$ \\
w & $\mathrm{Se}_{\mathrm{Mn}}$ & $-7$ & $+1158$ \\
w, stripe & $V_{\mathrm{Mn}}$ & $-3.06$ (unconstrained) & $0$ \\
w, stripe & $V_{\mathrm{Mn}}$ & $-5$ & $+213$ \\
w, stripe & $V_{\mathrm{Mn}}$ & $-7$ & $+447$ \\
\bottomrule
\end{tabular}
\par\smallskip\parbox{0.9\linewidth}{\footnotesize w: A-type order, 72-atom cell. w, stripe: stripe order, 64-atom cell of Table~\ref{tab:S11}.}
\end{table}

\begin{table}[h]
\centering
\caption{Kohn--Sham levels of the defect cells at 120/960~Ry (eV, relative to the valence-band maximum
of the host on the same $3\times3\times3$ grid, aligned through the Se-$4s$ centroid): Fermi level, highest level
below and lowest level above it (spin up/down), the magnitude of the smearing term $|TS|$ (meV), and the zero-width
correction to $E_f$, $|TS|/2$ (eV). The hosts have gaps of 1.51~eV ($\alpha$) and 1.71~eV (w) on this grid. Cells with a
constrained $M$ have one Fermi level per spin. In gapped cells $E_F$ only reflects where Gaussian neutrality places it.
These are Kohn--Sham levels, not transition levels.}
\label{tab:S13}
\begin{tabular}{llrrrrr}
\toprule
Phase & Defect & $E_F$ & HOMO & LUMO & $|TS|$ & $|TS|/2$ \\
\midrule
$\alpha$ & $V_{\mathrm{Mn}}$ & $-0.07$ & $-0.13$/$-0.09$ & $-0.06$/$-0.02$ & $110$ & $0.055$ \\
$\alpha$ & $V_{\mathrm{Se}}$ & $+1.23$ & $+0.86$/$+0.86$ & $+1.55$/$+1.55$ & $0$ & $0.000$ \\
$\alpha$ & $\mathrm{Se}_{\mathrm{Mn}}$ & $+0.75$ & $-0.01$/$+0.60$ & $+1.15$/$+0.83$ & $22$ & $0.011$ \\
$\alpha$ & $\mathrm{Mn}_{\mathrm{Se}}$ & -- & $+1.62$/$+1.61$ & $+1.68$/$+1.64$ & $138$ & $0.069$ \\
w & $V_{\mathrm{Mn}}$ & $-0.03$ & $-0.06$/$-0.03$ & $-0.01$/$+0.03$ & $132$ & $0.066$ \\
w & $V_{\mathrm{Se}}$ & $+1.22$ & $+0.18$/$-0.01$ & $+1.83$/$+1.80$ & $0$ & $0.000$ \\
w & $\mathrm{Se}_{\mathrm{Mn}}$ & $+1.06$ & $+0.38$/$+0.61$ & $+1.59$/$+1.52$ & $0$ & $0.000$ \\
w & $\mathrm{Mn}_{\mathrm{Se}}$ & -- & $+0.15$/$+1.65$ & $+1.90$/$+1.80$ & $39$ & $0.020$ \\
\bottomrule
\end{tabular}
\end{table}

\begin{table}[h]
\centering
\caption{Band 27 read off the unfolded spectrum in different ways, in meV, against the pristine curve at $u = 0.378$
(peak 232~meV): odd part at the pristine peak, and rms and largest deviation over the path outside the region
around $\Gamma$. Last column: the share of an artificial 100~meV offset between the spin references that appears in the result,
a measure of how far a Gaussian window centered on the reference pulls the result towards it.}
\label{tab:S14}
\begin{tabular}{llrrrr}
\toprule
Defect & Estimator & Peak & rms & Max & Leak \\
\midrule
$V_{\mathrm{Mn}}$ & Tracked peak, $\sigma = 0.03$~eV & $226$ & $27$ & $54$ & -- \\
$V_{\mathrm{Mn}}$ & Hard window, $\pm0.10$~eV & $230$ & $23$ & $80$ & -- \\
$V_{\mathrm{Mn}}$ & Hard window, $\pm0.15$~eV & $224$ & $20$ & $45$ & -- \\
$V_{\mathrm{Mn}}$ & Hard window, $\pm0.17$~eV & $222$ & $20$ & $38$ & -- \\
$V_{\mathrm{Mn}}$ & Gaussian window, $\tau = 0.06$~eV & $233$ & $19$ & $42$ & 37\% \\
$V_{\mathrm{Mn}}$ & Gaussian window, self-consistent center & $229$ & $23$ & $41$ & -- \\
$V_{\mathrm{Se}}$ & Tracked peak, $\sigma = 0.03$~eV & $235$ & $18$ & $43$ & -- \\
$V_{\mathrm{Se}}$ & Hard window, $\pm0.10$~eV & $229$ & $14$ & $34$ & -- \\
$V_{\mathrm{Se}}$ & Hard window, $\pm0.15$~eV & $228$ & $16$ & $36$ & -- \\
$V_{\mathrm{Se}}$ & Hard window, $\pm0.17$~eV & $234$ & $16$ & $40$ & -- \\
$V_{\mathrm{Se}}$ & Gaussian window, $\tau = 0.06$~eV & $233$ & $13$ & $31$ & 25\% \\
$V_{\mathrm{Se}}$ & Gaussian window, self-consistent center & $227$ & $16$ & $35$ & -- \\
$\mathrm{Se}_{\mathrm{Mn}}$ & Tracked peak, $\sigma = 0.03$~eV & $230$ & $32$ & $106$ & -- \\
$\mathrm{Se}_{\mathrm{Mn}}$ & Hard window, $\pm0.10$~eV & $234$ & $19$ & $49$ & -- \\
$\mathrm{Se}_{\mathrm{Mn}}$ & Hard window, $\pm0.15$~eV & $242$ & $24$ & $71$ & -- \\
$\mathrm{Se}_{\mathrm{Mn}}$ & Hard window, $\pm0.17$~eV & $234$ & $22$ & $69$ & -- \\
$\mathrm{Se}_{\mathrm{Mn}}$ & Gaussian window, $\tau = 0.06$~eV & $236$ & $17$ & $47$ & 30\% \\
$\mathrm{Se}_{\mathrm{Mn}}$ & Gaussian window, self-consistent center & $237$ & $26$ & $85$ & -- \\
$\mathrm{Mn}_{\mathrm{Se}}$ & Tracked peak, $\sigma = 0.03$~eV & $237$ & $45$ & $120$ & -- \\
$\mathrm{Mn}_{\mathrm{Se}}$ & Hard window, $\pm0.10$~eV & $229$ & $22$ & $55$ & -- \\
$\mathrm{Mn}_{\mathrm{Se}}$ & Hard window, $\pm0.15$~eV & $243$ & $16$ & $43$ & -- \\
$\mathrm{Mn}_{\mathrm{Se}}$ & Hard window, $\pm0.17$~eV & $272$ & $18$ & $45$ & -- \\
$\mathrm{Mn}_{\mathrm{Se}}$ & Gaussian window, $\tau = 0.06$~eV & $232$ & $19$ & $48$ & 41\% \\
$\mathrm{Mn}_{\mathrm{Se}}$ & Gaussian window, self-consistent center & $230$ & $37$ & $116$ & -- \\
\bottomrule
\end{tabular}
\par\smallskip\parbox{0.9\linewidth}{\footnotesize On the pristine cell the hard window is exact up to $\pm0.18$~eV and fails from $\pm0.19$~eV, where neighboring bands enter it, and the Gaussian window $\tau = 0.06$~eV is exact to 0.2~meV.}
\end{table}